\documentclass{aa}

\usepackage{graphicx}
\usepackage{txfonts,ulem}
\usepackage{mathrsfs}
\usepackage{color}
\usepackage{lscape}
\usepackage{xspace}
\usepackage{booktabs}
\usepackage{txfonts}
\usepackage{tabularx}
\usepackage{soul} 
\usepackage{makecell}

\newcommand{\kepler}{\textit{Kepler}\xspace}

\newcommand{\muHz}{\,\mu\mathrm{Hz}}
\newcommand{\teff}{\,T_{\mathrm{eff}}}
\newcommand{\logg}{\,\log{g}}
\newcommand{\Pcyc}{P_{\mathrm{cyc}}}
\newcommand{\Prot}{P_{\mathrm{rot}}}
\newcommand{\relI}{\Delta A (t)/ \overline{A}}
\newcommand{\relO}{\Delta \Omega (t)/ \overline{\Omega}}

\newcommand{\avrelI}{ \langle \Delta A(t)      / \overline{A}     \rangle}
\newcommand{\avrelO}{ \langle \Delta \Omega(t) / \overline{\Omega}\rangle}
\newcommand{\kicxample}{KIC002141150\xspace}
\newcommand{\logRHK}{\log{R\mathrm{'_{HK}}}}
\newcommand{\corr}{\langle r \rangle}

\begin{document}

   \title{Starspot rotation rates vs. activity cycle phase:\\ Butterfly diagrams of \kepler stars are unlike the Sun's}
 
    \author{M.~B.~Nielsen\inst{1}
        \and
        L.~Gizon \inst{2,3,1}
        \and
        R.~H.~Cameron \inst{2}
        \and
        M.~Miesch \inst{4}
		}

\institute{Center for Space Science, NYUAD Institute, New York University Abu Dhabi, PO Box 129188, Abu Dhabi, UAE\\
           \email{mbn4@nyu.edu}
           \and 
           Max-Planck-Institut f{\"u}r Sonnensystemforschung, Justus-von-Liebig-Weg 3, 37077 G{\"o}ttingen, Germany
           \and 
           Institut f{\"u}r Astrophysik, Georg-August-Universit{\"a}t G{\"o}ttingen, 
           Friedrich-Hund-Platz 1, 37077 G{\"o}ttingen, Germany
           \and 
           High Altitude Observatory, National Center for Atmospheric Research, Boulder, CO 80307-3000, USA}

\date{Received XXXXXX XX, 2018; accepted XXXXXX XX, 2018}

  \abstract
   {During the solar magnetic activity cycle the emergence latitudes of sunspots change, leading to the well-known butterfly diagram. This phenomenon is poorly understood for other stars since starspot latitudes are generally unknown. The related changes in starspot rotation rates caused by latitudinal differential rotation can however be measured.}
   {Using the set of 3093 \kepler stars with activity cycles identified by \citet{Reinhold2017}, we aim to study the temporal change in starspot rotation rates over magnetic activity cycles, and how this relates to the activity level, the mean rotation rate of the star, and its effective temperature.}
   {We measure the photometric variability as a proxy for the magnetic activity and the spot rotation rate in each quarter over the duration of the \kepler mission. We phase-fold these measurements with the cycle period. To reduce random errors we perform averages over stars with comparable mean rotation rates and effective temperature at fixed activity-cycle phases.}
   {We detect a clear correlation between the variation of activity level and the variation of the starspot rotation rate. The sign and amplitude of this correlation depends on the mean stellar rotation and -- to a lesser extent -- on the effective temperature. For slowly rotating stars (rotation periods between $15-28$ days) the starspot rotation rates are clearly anti-correlated with the level of activity during the activity cycles. A transition is observed around rotation periods of $10-15$ days, where stars with effective temperature above $4200$K instead show positive correlation.}
   {Our measurements can be interpreted in terms of a stellar ``butterfly diagram'', but these appear different from the Sun's since the starspot rotation rates are either in phase or anti-phase with the activity level. Alternatively, the activity cycle periods observed by \kepler are short (around $2.5$ years) and may therefore be secondary cycles, perhaps analogous to the solar quasi-biennial oscillations.}
   \titlerunning{Starspot rotation rates {\it vs.} activity cycle phase}
   \authorrunning{Nielsen et al.}

   \maketitle

\section{Introduction}
\label{sec:intro}
On the Sun, the latitudinal migration of sunspots produces the well-known butterfly diagram \citep{Maunder1904}. As the solar magnetic activity cycle progresses, the spot coverage of the Sun increases, while the typical emergence latitude moves closer to the equator. After activity maximum the spots continue to migrate further toward the equator, eventually stopping at approximately $8^{\circ}$ latitude, at which point the next 11 year cycle begins again with spots appearing at around $30^{\circ}$ latitudes. The emergence latitudes covered by sunspots span a range of rotation rates, due to the latitudinal differential rotation of the Sun. Observations of spots and active regions are therefore among the methods that have been used to establish the solar latitudinal differential rotation profile \citep[see, e.g., the review by][]{Beck2000}. The differential rotation profile, combined with the butterfly diagram are clear observable features of the solar magnetic dynamo, and their characteristics therefore act as strong constraints on dynamo models \citep[see, e.g., reviews by][]{Rempel2008,Charbonneau2010}.

Other stars also exhibit magnetic activity, ranging from young fast rotators with irregular magnetic activity through older, slower, rotators that show smooth periodic cycles like the Sun, to in some cases not showing any long term variability at all \citep{Wilson1978,Baliunas1995,Hall2009}. These activity cycles are coupled with a brightening or dimming of the star. Relatively old and slow rotators like the Sun show a brightening of $\sim0.1\%$ along with increased photometric variability during activity maximum \citep{Froehlich1987,Hall2007}. This is in spite of an overall increase in the number of dark spots on the stellar surface, and is attributable to a similarly enhanced number of faculae and plage regions \citep{Foukal1979}. However, previous studies \citep{Radick1998,Hall2009} found that younger and more active stars than the Sun predominantly tend to show the inverse behavior, namely a dimming during activity maximum rather than a brightening. \citet{Radick1998} suggested that this change is due to younger, active stars exhibiting spot-dominated variability while older, less active stars eventually become faculae dominated. \citet{Montet2017} studied photometric variability in 463 solar-like stars (in terms of $\logg$ and $\teff$) and detected a transition from spot dominated (dimming during variability maximum) to faculae dominated (brightening during variability maximum) photometric variability at a rotation period of $\sim 10-20$ days. This falls neatly in line with efforts to model the variability of the total solar irradiance (TSI), which show that the Sun is predominantly faculae dominated \citep[][]{Chapman1987,Shapiro2016}.

As with the Sun, the spot rotation rate of stars can also be expected to vary over the course of their respective activity cycles. While still challenging, measurements of latitudinal differential rotation have now been performed on a variety of other stars. The methods employed are wide ranging and include: the study of power spectra of photometric light curves \citep[e.g.][]{Lanza1993,Reinhold2013a,Reinhold2015}, time variations of spectral line profiles with Doppler imaging \citep{Donati1997,Collier2002,Barnes2001} and Zeeman Doppler imaging \citep{Petit2004,Barnes2005}, changes in starspot rotation rates \citep{Henry1995,Messina2002}, variability in the magnetically sensitive Ca HK lines \citep[e.g.][]{Donahue1996}, and more recently using asteroseismology \citep{Benomar2018}. However, since activity cycles tend to have periods of several years \citep{Baliunas1995}, only a few studies have been able to trace rotation periods over the course of one or more cycles \citep{Donahue1992,Messina2003}. 

In this work we extend the sample of stars with both measured activity cycles and rotation periods to include a recently published activity cycle catalog by \citet{Reinhold2017}. We then study the change in the spot rotation rate over time and how it relates to the activity cycles of the stars. 

\section{Stellar activity cycles from Kepler data}
\label{sec:sample}
\begin{figure}
   \centering
   \includegraphics[width=\hsize]{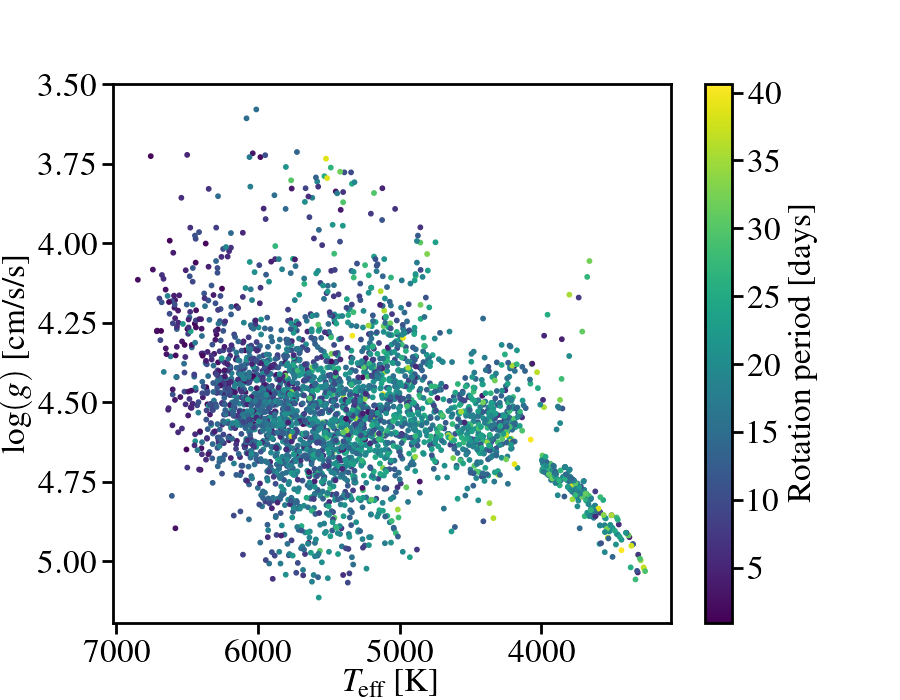}
      \caption{Effective temperature $\teff$ and surface gravity $\logg$ of the stars with measured activity cycle periods from \citet{Reinhold2017}. The color code denotes rotation periods as measured by \citet{McQuillan2014}.}
         \label{fig:hrd}
\end{figure}

\citet{Reinhold2017} use the photometric variability of \kepler light curves as a proxy for magnetic activity. This method is motivated by solar observations showing increased variance of the measured flux during solar activity maximum, and conversely a decrease during activity minimum. In the Sun this enhanced variability stems from an increase in the number of dark spots, bright faculae, and plage regions and their evolution over time. The assumption is that other stars show a similar cyclic change in photometric variability through the activity cycle.

The photometric variability used by \citet{Reinhold2017} was originally defined by \citet{Basri2011} as the $5$th to $95$th percentile range of the relative flux, and is computed for each quarter ($\sim90$ days) of \kepler observations over the course of the mission lifetime ($\sim3.4$ years). \citet{Reinhold2017} fit a sine function to the variability measurements, which was tested against a $5\%$ false alarm probability. The catalog was based on the rotation catalog by \citet{McQuillan2014} which contains rotation periods of 34030 stars. Out of these \citet{Reinhold2017} find a total of $3203$ stars that show significant periodicity of their photometric variability. The measured period of the sine function is then assumed to be the period of the activity cycle. The measured cycle periods range from $0.5-6$ years, with a median of $\sim3$ years. 

Figure~\ref{fig:hrd} shows the effective temperature $\teff$ and surface gravity $\logg$ for these $3203$ stars (adapted from \citet{Huber2014}, along with their rotation periods from \citet{McQuillan2014}. The sample predominantly consists of cool main-sequence stars, with a range of rotation periods spanning $1-41$ days, and a median of $\Prot\approx16$ days; the temperature range is from approximately $3200$K to $6900$K with a median of $\teff \approx 5500$K. Based on the trend of slower rotation with increasing age \citep[see, e.g.,][]{Barnes2010}, this sample of stars is likely on average slightly younger than the Sun. 

An inspection of the cycle periods revealed spuriously high numbers of stars with periods of either one year or $185$~days. We checked the light curves of all stars with these cycle periods. For stars with $\Pcyc \approx 1$ year the light curves often show discontinuous jumps in the relative flux amplitude every $4$ quarters. This is most likely caused by the quarterly roll of the \kepler spacecraft. Every $4$ quarters any given star will land on the same CCD, thereby being subject to the same instrumental noise sources, thus inducing a $1$ year periodicity in the variability measurements. We removed stars that showed this clear discontinuity in the flux variability, but retained stars that showed a smooth transition between quarters. For stars with $\Pcyc \approx 185$ days (77 stars) the cause is less certain, but it may also be related to instrumental effects since it constitutes approximately 2 quarters of observations. An inspection of the light curves did not reveal any remarkable differences compared to the stars in the remainder of the sample. However, we find it unlikely that such a spuriously large number of stars have the same cycle period, and therefore opt to remove them from the sample entirely. In total 110 stars were removed from the sample. This leaves 3093 stars for further analysis.

It should be noted that this sample is likely biased towards stars with significant and coherent photometric variability, as this forms the basis for the rotation measurement method employed by \citet{McQuillan2014}. \citet{Montet2017} suggest that stars with rotation periods shorter than $\Prot \approx 20$ are predominantly spot dominated. While this period range constitutes the majority of stars in our sample, we only find an overlap of $83$ stars with \citet{Montet2017}. Of these, $80$ stars were defined as spot dominated. Since this only represents a fraction of our total sample we cannot say conclusively if the majority of stars in our sample are spot- or faculae-dominated. 

In addition to the above bias, these are all stars with cyclic activity levels. Young, fast rotating stars with strong but incoherently varying activity or old, inactive stars with no variation at all, are therefore underrepresented. The range of cycle periods is between $0.5$ and $6$ years, so exact analogs of the solar activity cycle are also not included. 

Searching the catalog of $\logRHK$ values collated by \citet{Karoff2016} we find 95 stars overlapping with our sample. These stars have an average $\logRHK = -4.569$, while the Sun shows a significantly lower $\logRHK = -4.895$. Again, this is not a sufficient overlap to draw conclusions about the average activity level of our sample. However, considering the high amplitude and coherent variability in the majority of our sample, we suggest that these stars are likely on average more active than the Sun.  

\section{Measuring stellar photometric variability and starspot rotation rate in each quarter}
\label{sec:measure}
\begin{figure}
   \centering
   \includegraphics[width=\hsize]{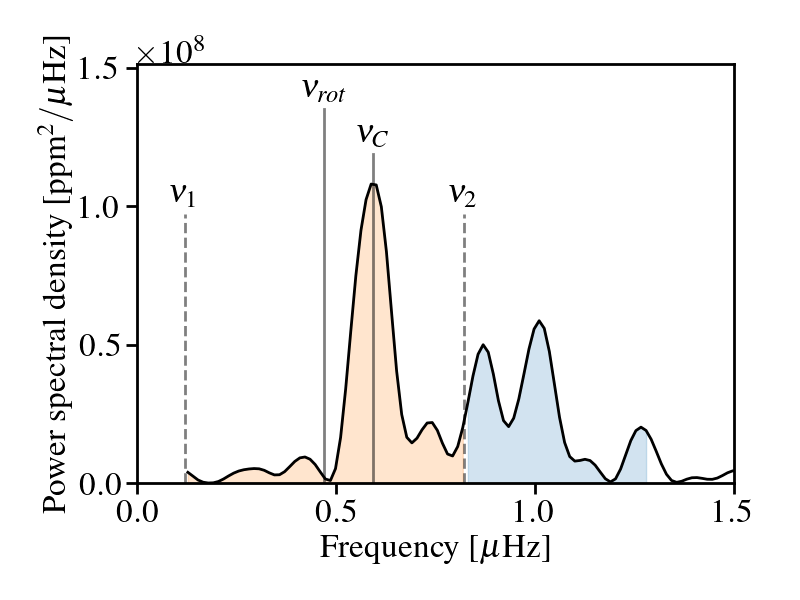}
      \caption{Illustration of the measured variability quantities in an example power spectrum of KIC002141150 during Q14. We use the interval between $\nu_1$ and $\nu_2$ (orange shaded region) as a representation of the starspot variability. The variance in the light curve caused by active regions is proportional to the integral of the power in this interval. The rotation rate at the spot latitude is measured by the centroid $\nu_C$. The interval is centered on the average rotation rate $\nu_{rot}$ of the star, obtained from \citet{McQuillan2014}. The blue region shows the first harmonic of the rotation period.}
         \label{fig:concept}
\end{figure}

In this work we focus on measuring the amplitude and characteristic frequency of variability induced by magnetic activity, and how it varies over time with the activity cycle. These quantities are measured from the power spectrum\footnote{Computed using the Lomb-Scargle method \citep{Lomb1976,Scargle1982}} of the time series of each star from each observation quarter. These measurements are then phase-folded using the cycle period and phase values obtained from \citet{Reinhold2017}, to study average changes in spot rotation rate and activity level (see  Section~\ref{sec:var}). 

We use the long-cadence \kepler light curves from all quarters of the mission. We use the PDC-msMAP pre-processed version of the light curves to avoid as much contamination of systematic variability as possible. Note that many of the stars in the sample have rotation periods longer than 20 days, beyond which the PDC-msMAP pipeline begins to significantly reduce the amplitude of intrinsic stellar variability \citep{Gilliland2015}.

For each star we start by defining a frequency range around the known average rotation rate $\nu_{rot}$, bounded by $\nu_{1} = \nu_{rot}-\delta\nu$  and $\nu_{2} = \nu_{rot} + \delta\nu$, where $\nu_{rot}$ is taken from \citet{McQuillan2014} (shown in Fig.~\ref{fig:concept}). The half-width $\delta\nu$ of the frequency range is defined as
\begin{equation}
\delta \nu  = \left\{ {\begin{array}{*{20}c}
   \alpha         & \rm{for}\quad0.5\,\nu _{rot}  < \alpha  \\
   0.5\nu _{rot}  & \rm{for}\quad0.5\,\nu_{rot}  \ge \alpha   \\
\end{array}} \right.
\end{equation}

For slow rotators, this definition of $\delta\nu$ minimizes the potential contribution of power from the 1st harmonic of the rotation rate. The value of $\alpha$ is set to $0.4\muHz$, which is chosen in order to allow for variation of the rotation rate, while also reducing the contribution from instrumental variability on time-scales longer than the stellar rotation period.

We define the spot rotation rate as $\Omega = 2\pi\nu_C$. Here $\nu_C$ is the centroid, which is estimated by the power-weighted sum of the frequency bins in the range from $\nu_{1}$ to $\nu_{2}$. For a single monolithic peak in the power spectrum (corresponding to a single, large spot or group of spots) the centroid will lie close to the frequency of maximum power of the peak. For cases where multiple frequencies show increased power (corresponding to spots at multiple latitudes) the centroid will represent a weighted average frequency of the variability. For stars with strong differential rotation, spots appearing at various latitudes will cause $\nu_{C}$ to vary around $\nu_{rot}$ in accordance with the spot emergence pattern.

The integral of the power $\mathcal{P}$ in this frequency range is taken as a proxy for the amplitude $A$ of the spot variability, such that $A = \left[ {\int_{\nu _{1} }^{\nu _{2} } {\mathcal{P}\,d\nu } } \right]^{1/2}.$ This is similar to the photometric range used by \citet{Reinhold2017}. However, unless properly filtered in frequency the photometric range is sensitive to additional sources of variability, such as granulation or shot noise. Here, $A$ is akin to the band-pass filtered photometric range. By confining the frequency range to a narrow region around the rotation rate, we ensure that the majority of the variability that contributes to $A$ stems from the rotation modulation. 

We perform these measurement for each individual quarter, such that for each star we obtain $A(t_q)$ and $\Omega(t_q)$. Where $t_q$ is the median time of observation of each quarter $q$. For simplicity we drop the subscript $q$ in the following. The activity level $A(t)$ and $\Omega(t)$ are measured relative to their average values across all quarters, $\overline{A}$ and $\overline{\Omega}$ respectively. This gives us the relative change over time in photometric variability, $\relI$, and spot rotation rate, $\relO$.

\begin{figure*}
   \centering
   \includegraphics[width=\hsize]{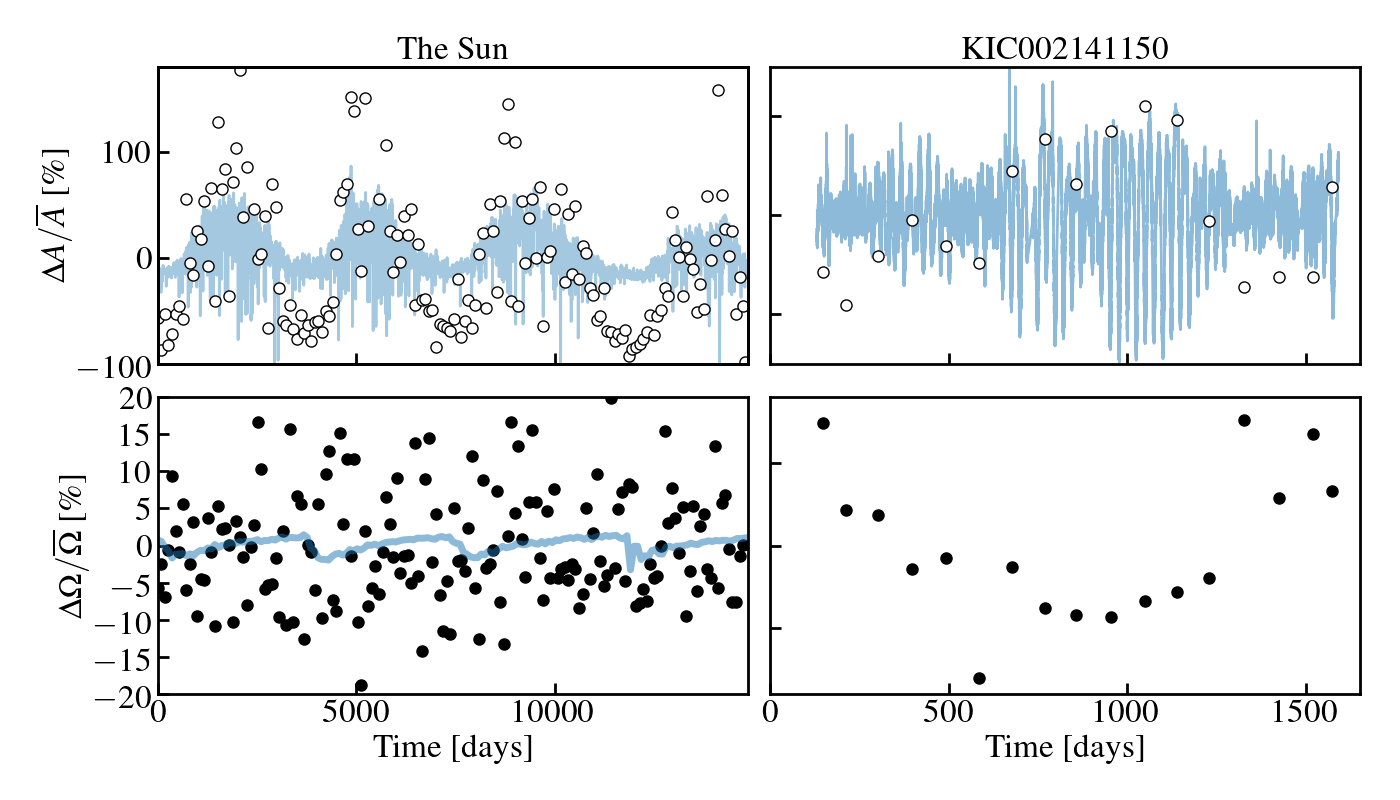}
      \caption{Activity variability amplitude $\relI$ (top, open circles) and relative spot rotation rate $\relO$ (bottom, filled circles) for the Sun (left) and an example star \kicxample (right), which has a cycle period $\Pcyc \approx 4$ years. The solar values are based on TSI observations covering approximately $42$ years (blue in top left frame). The \kepler data (blue in top right frame) consist of quarterly measurements and so the solar data are likewise separated into $90$ day bins (equivalent to \kepler quarters). In the lower left frame the blue curve indicates the solar $\relO$ computed based on the latitudes of directly observed sunspots.}
         \label{fig:cycle_example}
\end{figure*}

Figure~\ref{fig:cycle_example} shows $\relI$ and $\relO$ for the Sun and one of the stars in the sample: \kicxample, which has a cycle period of $\Pcyc \approx 4$ years. The solar values are measurements of a composite set of TSI\footnote{Downloaded from \url{ftp://ftp.pmodwrc.ch/pub/Claus/ISSI_WS2005/ISSI2005a_CF.pdf}. PMOD/WRC composite v. 42\_65\_1709} measurements from multiple spacebased radiometers \citep{Froehlich2006}. These observations span almost $42$ years, covering cycles $21$ through $24$. 

The variance of the TSI is enhanced by variability from magnetic activity during activity maximum, and thus the integral $A$ that we are measuring is similarly enhanced. Using the Sun as a reference, in the following we therefore define the maximum of $\relI$ as the activity maximum and likewise for the activity minimum. However, it should be noted that we have no information concerning the variation of the chromospheric emission and other magnetic activity indices for the stars in our sample, and so this definition is only based on photometric variability. 

For both the Sun and \kicxample $\relI$ shows a variation with the cycle period. On the other hand the spot rotation rate $\relO$ measurements for the Sun do not show any apparent variation. Sunspots tend to have lifetimes comparable the rotation period of the Sun \citep[e.g.,][]{Petrovay1997}, which makes measuring the solar rotation rate from the TSI power spectrum very difficult. In addition, the scatter of spot emergence latitudes during the solar cycle spans up to $\approx10$ degrees \citep{Hathaway2010}. This, combined with the short lifetimes makes measuring an average rotation rate challenging. Similarly, measuring the solar differential rotation from the TSI power spectrum is also challenging. In the following we therefore use the relative rotation rate of directly observed sunspots during the same period (shown in blue in the lower left frame of \ref{fig:cycle_example}). These are computed from sunspot latitudes\footnote{RGO and NOAA measurements downloaded from https://solarscience.msfc.nasa.gov/greenwch.shtml} and using the solar rotation profile by \citet{Snodgrass1990}. In contrast to the centroid values the variation of the directly observed sunspot rotation rates clearly vary with the activity cycle of the Sun.

For \kicxample, however, the $\relO$ measurements do show variability that is on similar timescales as the cycle period, and anti-correlated with $\relI$, indicating latitudinal differential rotation. Here it is worth emphasizing that other stars in this sample do not necessarily show the same clear trends in $\relO$ as in $\relI$. Indeed, measuring stellar differential rotation using only photometric measurements is often difficult, mainly due the evolution of the spot signal as the spot grows and decays \citep[see, e.g.,][]{Aigrain2015}. In the power spectrum the effect of spot evolution is broadening of the distribution of power around the rotation peak. The lifetime of sunspots has indeed been shown to be correlated with the activity cycle \citep{Henwood2010}, however, when averaging over multiple stars we expect this effect to be symmetric around the rotation peak and will therefore not have a systematic effect on $\relO$.  

\section{Starspot rotation rate versus activity cycle phase}
\label{sec:var}
\begin{figure*}
   \centering
   \includegraphics[width=0.49\hsize]{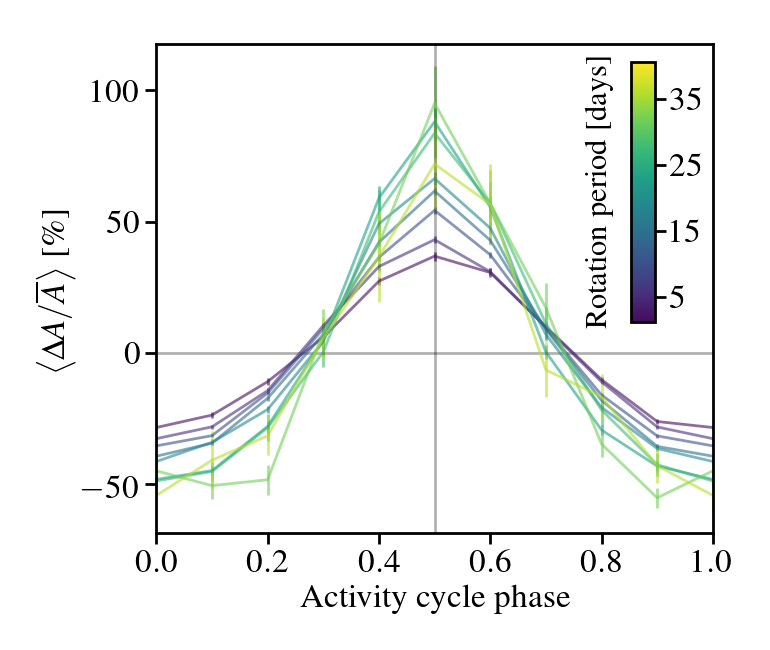}
   \includegraphics[width=0.49\hsize]{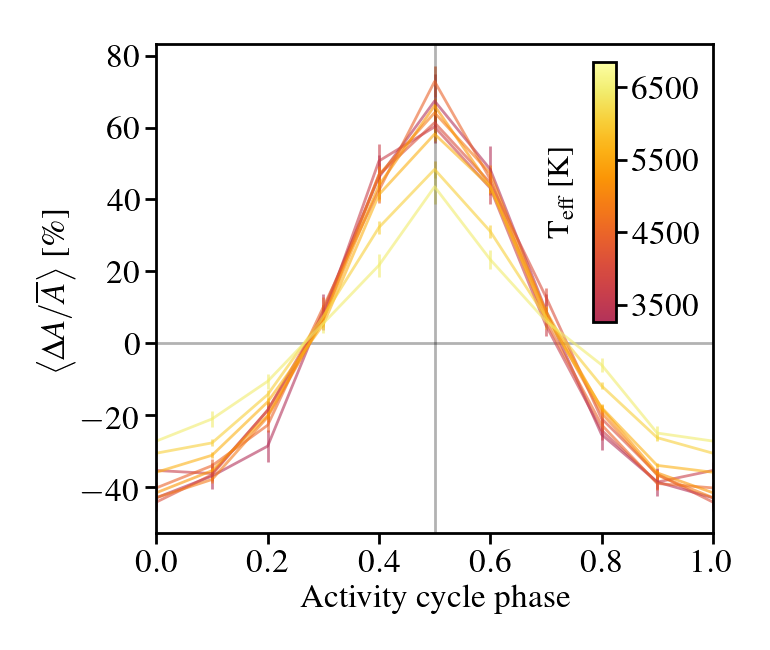}
   \includegraphics[width=0.49\hsize]{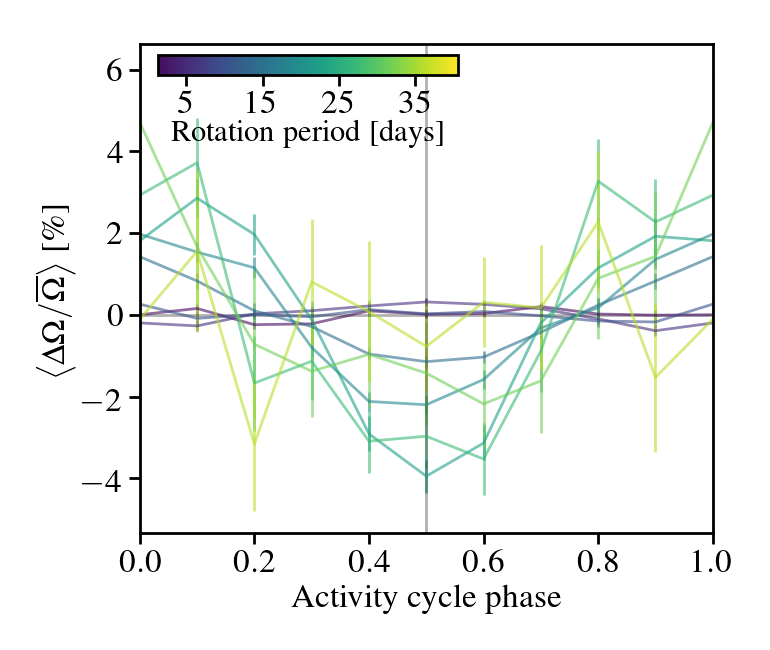}
   \includegraphics[width=0.49\hsize]{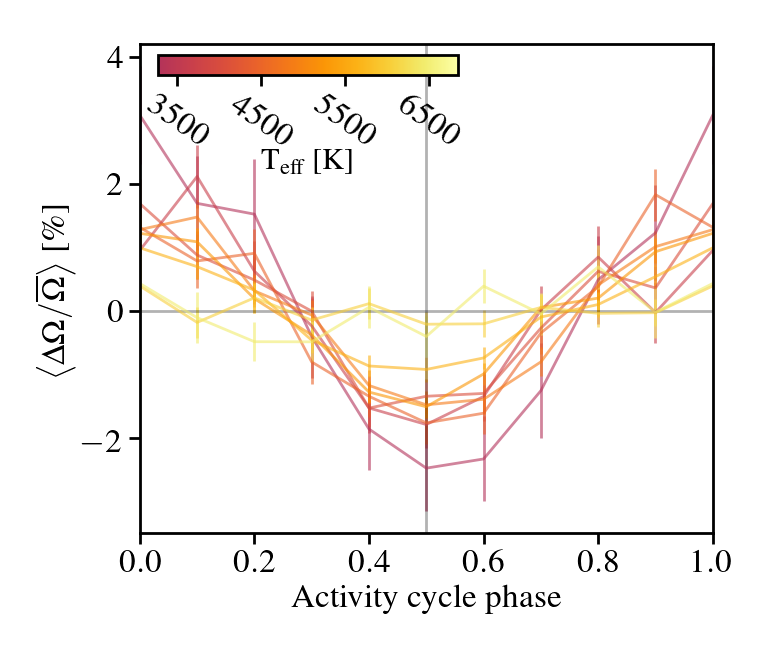}
   \caption{Top row: The averaged photometric variability $\avrelI$ as a function of phase for stars with different rotation periods (left) and effective surface temperatures, $\teff$ (right). Bottom row: Same as the top row, but for $\avrelO$ instead. The stars are binned by period and effective temperature, with bin widths of $\approx 4$ days and $\approx 400$K respectively. For clarity we also show $\avrelI=0$ and $\avrelO=0$ and the half-way point $\phi = 0.5$ of the activity cycle in gray.}
         \label{fig:var}
\end{figure*}

While attempting to measure the variation of the spot rotation rate with the activity cycle for individual stars is typically not feasible, especially if the spot lifetime is short, it is possible to do so when averaging the $\relO$ measurements of multiple stars. We first phase-fold the measurements\footnote{These will be made available as online material at the CDS.} of all the stars using the cycle periods $\Pcyc$ and phases $\phi_0$ from \citep[from][]{Reinhold2017}, and then compute the averages $\avrelI$ and $\avrelO$ in each phase bin. Here $\langle\rangle$ represents an average of quarterly measurements of multiple stars at the same phase of their respective activity cycles. Figure~\ref{fig:var} shows $\avrelI$ and $\avrelO$ as a function of activity cycle phase, where the errors represent the standard errors on the mean. The bin widths in rotation period and effective temperature are $\approx4$ days and $\approx400$K respectively.

The mean rotation rate $\overline{\Omega}$ and the effective surface temperature, $\teff$, are thought to be two of the important parameters that influence the dynamo mechanism \citep[see, e.g.,]{Charbonneau2010}. In Fig.~\ref{fig:var} we have therefore divided the measurements of $\avrelI$ and $\avrelO$ into separate rotation and temperature bins, and show the variation of these parameters with cycle phase for each bin. 

\subsection{Shape of activity cycles}
\label{sec:photvar}
As expected from Fig.~\ref{fig:cycle_example} the values of $\avrelI$ clearly vary with the activity cycle, with generally a sharp peak followed by a shallow minimum. Note that by the construction of the catalog by \citet{Reinhold2017} the general shape of the variation cannot change dramatically, as the detection of the cycles were made using sine fits to the variability. A strong deviation from a symmetric profile might produce a false alarm probability greater than the $5\%$ imposed by \citet{Reinhold2017}, and thus would have been removed from the sample. Note that for clarity in the following sections we have phase shifted our measurements of $\avrelI$ and $\avrelO$ compared to \citet{Reinhold2017}, such that the minimum of $\avrelI$ (for all the stars in the sample) is at phase $\phi = 0$.

It is clear that the amplitude of $\avrelI$ changes with the mean rotation period, and only marginally so for changes in temperature. For the fast rotators the change in $\avrelI$ over the cycle is much smoother, and lower amplitude. In contrast, the slow rotators have a more peaked structure during activity maximum, with an amplitude relative to the mean of about a factor of 2-3 greater than the fast rotators. Compared to the fast rotators, the slow rotators also appear to have more pronounced excursions to high levels of variability at cycle maximum, whereas the levels of variability at cycle minimum in both cases remain similar. 

Temperature appears to have a smaller impact on the variation of the photometric variability. Only the very hottest stars in the sample show any notable difference in $\avrelI$. These stars are on average also faster rotators with a mean rotation period of $\approx 10$ day, and so the effect of averaging over many stars with less coherent variability could be the cause of the reduced amplitude that is seen here.  

\subsection{Detection of changes in starspot rotation rates with cycle phase}
\label{sec:rotvar}
The change in $\avrelO$ appears to vary more dramatically with the mean rotation rate and temperature, than for $\avrelI$. The slow and cool stars show the largest variation in $\avrelO$, while the fast and hot stars show very little. The amplitude of $\avrelO$ for slow rotators is comparable to that of the Sun, but has a more sinusoidal form as opposed to the sharp decrease and slow increase seen for sunspot rotation rates. In general, the correlation between the $\avrelO$ and the activity cycle is negative, that is, the observed spot rotation rate decreases around activity maximum and vice versa during activity minimum. This is different from the variation seen on the Sun (see Fig.~\ref{fig:cycle_example}).

It is important to note that the mean rotation period and the effective surface temperature are strongly correlated \citep[see e.g.,][]{Nielsen2013,McQuillan2014}, with cool stars being on average slower rotators. The observed change in $\avrelO$ with increasing temperature is therefore likely tied to the change in rotation period (see below). 

\label{sec:cor}
\begin{figure*}
   \centering
   \includegraphics[width=\hsize]{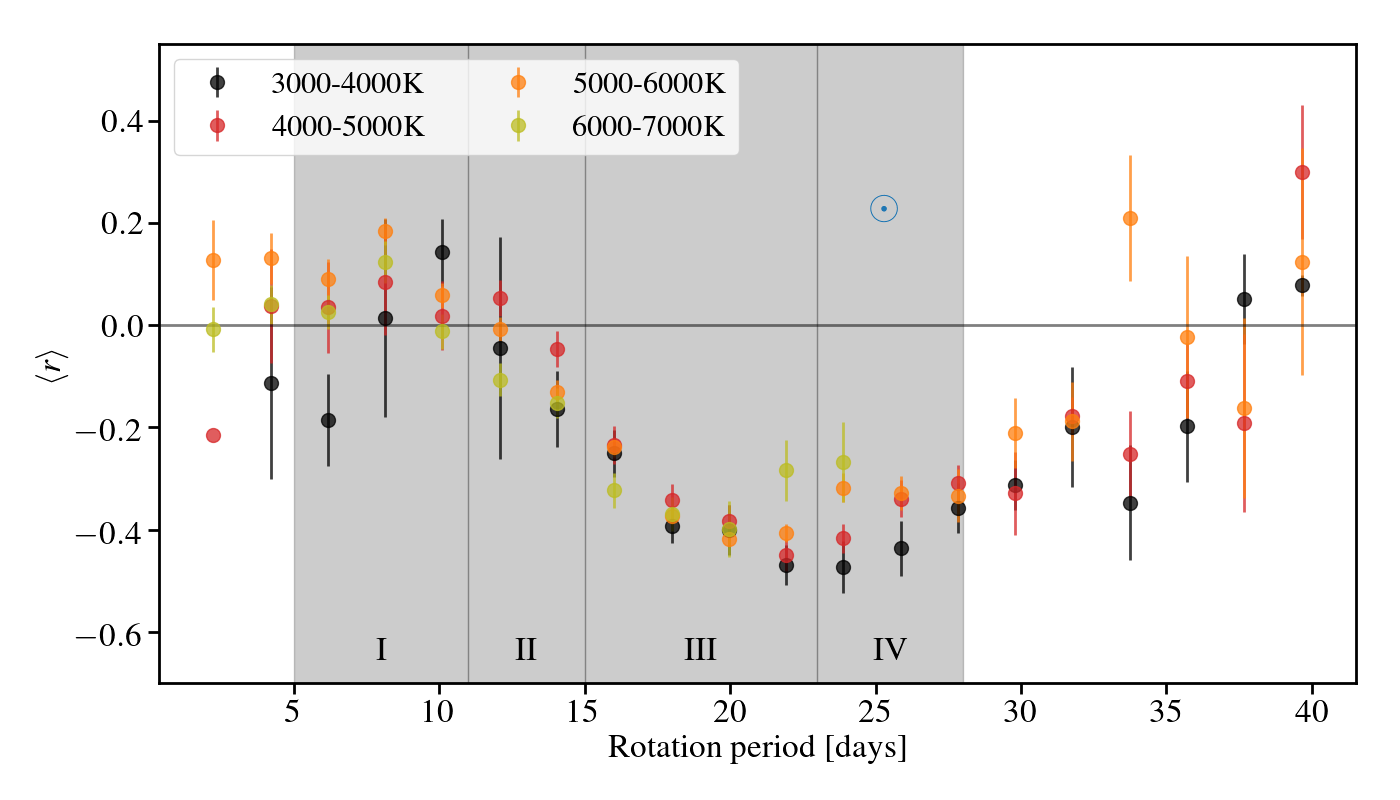}
      \caption{Average correlation coefficient $\corr$ of $\relI$ and $\relO$ as a function of rotation period. Colors denote sub-samples of different temperatures. Period ranges of particular interest are shaded gray and labeled in roman numerals. Note that the number density diminishes to only a few tens of stars at very short and very long rotation periods, particularly for the extremes of the temperature range. The sample does not contain any stars with temperatures above $6000$K with rotation periods longer than $25$ days. The correlation of the solar spot rotation rates (blue in Fig.~\ref{fig:cycle_example}) and variability amplitude is indicated by a blue $\odot$ at the Carrington rotation period. 
      } 
         \label{fig:corr}
\end{figure*}

\subsection{Average correlation between $\relI$ and $\relO$}
While the values of $\avrelI$ indicate that a cycle is present at all temperatures and mean rotation periods, there is a stark change in the variation of $\avrelO$ as a function of cycle phase. To investigate this in more detail for our sample, we show in Fig.~\ref{fig:corr} the averaged Pearson correlation coefficient $\corr$ of $\relI$ and $\relO$, for stars of different temperature and average rotation periods. We divide the period range into four ranges of particular interest, which we will focus on in following. 

\begin{enumerate}[I]
\item{$\Prot = 5$-$11$ days: These stars predominantly show very little correlation between the activity cycle and changes in the spot rotation rate. However, a weak temperature dependence is visible. The hotter ($\teff > 5000\mathrm{K}$) stars show a marginal positive correlation on average, while the cool stars are closer to $\corr = 0$ or marginally negative.}
\item{$\Prot = 11$-$15$ days: The stars in this range appear to transition from $\corr \approx 0$ toward negative correlation on average.}
\item{$\Prot = 15$-$23$ days: These stars begin to show the same anti-correlation as the slower rotators. However, there is very little scatter due to differences in effective temperatures, compared to the faster rotators.}
\item{$\Prot = 23$-$28$ days: Stars in this period range have similar rotation rates as the Sun, but also show the largest degree of anti-correlation of all the stars in our sample. Similar to the fast rotators, $\avrelO$ and $\avrelI$ for the hotter stars in this period range appear to be somewhat less anti-correlated than their cooler counterparts. As was seen in Fig.~\ref{fig:var} the hot stars have slightly lower variability in $\avrelI$. This may be caused by less coherent activity levels, which may explain the observed decorrelation. However, we note that there are only very few hot stars with these periods (there are no hot stars with periods longer than 25 days in our sample).}
\end{enumerate}

Stars with rotation periods longer than $\sim30$ days appear to show a reduced correlation with the activity cycle, however the sample density in this period range is very low (only $86$ stars). Furthermore, the frequency window used to measure $\Omega$ reaches $\nu=0\,\muHz$ at $\nu_{rot}=0.4\,\muHz$, corresponding to $\Prot \approx 29$ days. For rotation periods longer than this the measurements of the centroid may be biased if the latitudinal differential rotation is strong. It is therefore unclear if the upward trend starting at $\overline{P} =30$ days is real. 

For reference we compute the corresponding correlation coefficient using the spot rotation rates from directly observed sunspots at various latitudes (blue curve in Fig.~\ref{fig:cycle_example}) over the 11 year solar cycle. The correlation coefficient is $\langle r \rangle \approx 0.22$. The reason for the positive correlation in the Sun is due to the migration of sunspot during the solar cycle. Near the start of the cycle sunspots initially appear at high latitudes. Following this the activity rises quickly, which leads to a positive correlation in the first phase of the cycle. Post-maximum, the activity level decreases while the spots continue to migrate toward faster rotating latitudes at toward the equator, thus leading to a negative correlation in the latter part of the cycle.

\subsection{Relationship between the averages $\avrelI$ and $\avrelO$}
\begin{figure*}
   \centering
   \includegraphics[width=0.95\hsize]{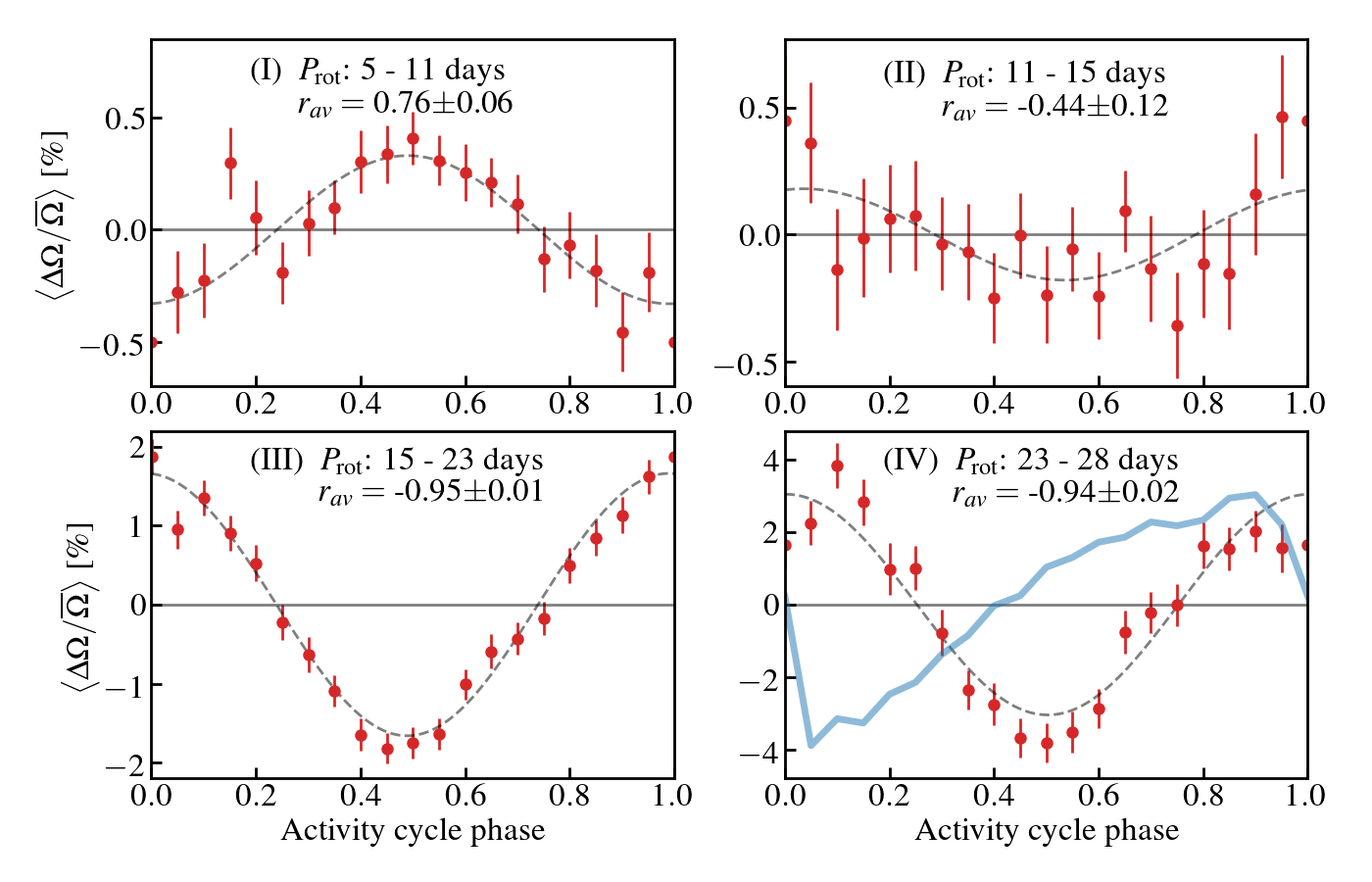}
      \caption{Averages of starspot rotation rates (red) over the stars in each sample I through IV. For clarity in frame I we only show stars with $\teff > 4200$K; stars cooler than this show similar anti-correlated behavior as slower rotators. In the remaining frames the entire sample in the respective period range is represented. In frame IV we also show in blue the variation of the sunspot rotation rate multiplied by a factor of $2.5$ for comparison. By construction, the activity maximum occurs close to phase $\phi = 0.5$ (see Fig.~\ref{fig:var}). $r_{av}$ is the correlation between the $\avrelI$ and $\avrelO$ of the stars in each period range. The dashed curves are sinusoidal fits to the averages $\avrelO$, whose parameters are provided in Table \ref{tab:chars}.}
         \label{fig:flip}
\end{figure*}

\begin{figure}
   \centering
   \includegraphics[width=\hsize]{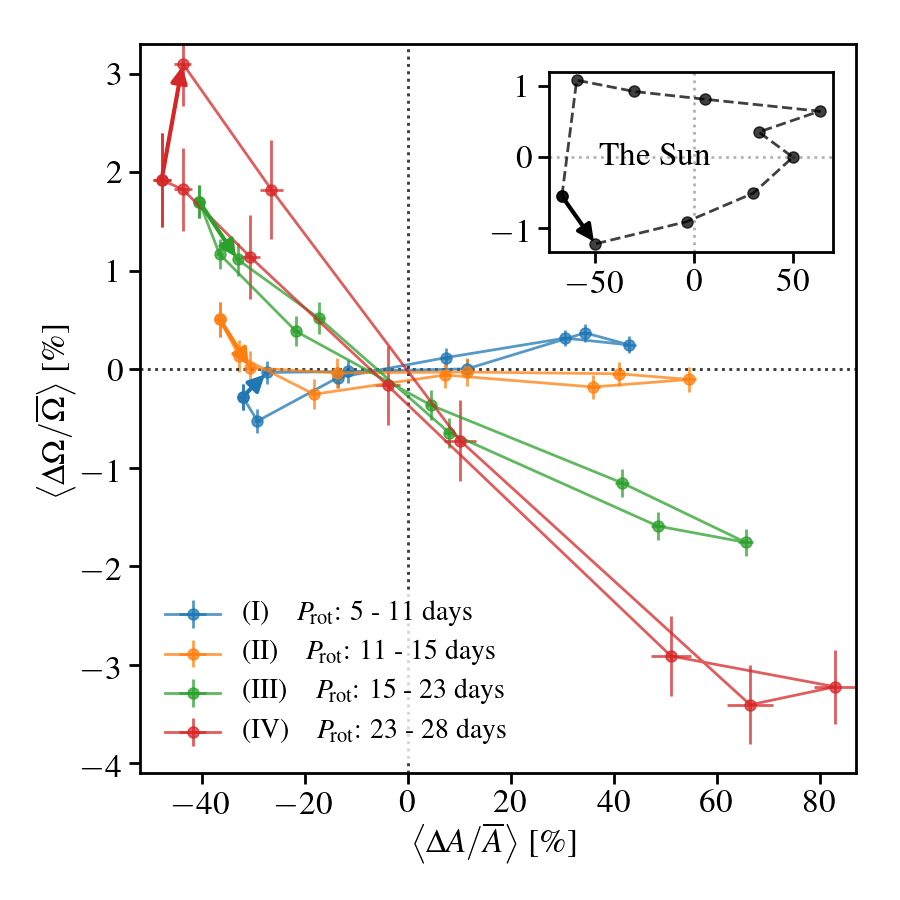}
      \caption{$\avrelO$ as a function of $\avrelI$ for the four period ranges of interest. For stars with $\Prot$ between 5 and 11 days (blue) only stars with $\teff > 4200$K are shown for clarity. The inset shows the corresponding values for the Sun. Arrows denote the progression of the cycle phase, starting at activity minimum. Dotted lines indicate the position of the origin.}
         \label{fig:eliptical}
\end{figure}

Figure~\ref{fig:flip} shows the change in $\avrelO$ as a function of cycle phase, while Fig.~\ref{fig:eliptical} shows it as a function of $\avrelI$. Here we also show the correlation of the averages, $r_{av}$, between $\avrelI$ and $\avrelO$. Note that in Fig.~\ref{fig:flip} the activity maximum occurs at phase $\phi\approx 0.5$ for all the period ranges (see Fig.~\ref{fig:var}). These figures will be discussed in the following. 

\begin{enumerate}[I]
\item $\Prot = 5$-$11$ days: This period range is represented in blue in Fig.~\ref{fig:eliptical}, and the top left frame of Fig.~\ref{fig:flip}. The correlation coefficients in this period range are temperature dependent, and so for clarity in the two figures we omit stars cooler than $4200$K, as the variation of $\avrelO$ for stars cooler than this is the same as in the remaining period ranges. The positive correlation seen for the hotter stars in this period range is clearly visible in Fig.~\ref{fig:flip}, and appear roughly sinusoidal (indicated by the dashed gray sine-curve). 

\item $\Prot = 11$-$15$ days: On average, $\avrelO$ in this period range does not vary substantially and shows little if any correlation with the activity cycle. Dividing this period range into narrower period bins shows no difference between the stars with periods around $11$ days compared to $15$ days. The decrease in the amplitude of $\avrelO$ is therefore not likely due to averaging two populations in terms of period, with positive or negative correlation respectively. These stars still exhibit a variation in photometric variability, indicating an activity cycle is present.  

\item $\Prot = 15$-$20$ days: These stars begin to show the same anti-correlated profile for $\avrelO$ as is seen in the slower rotators. However, $\avrelO$ appears have a more sinusoidal variation as indicated by the sine fit. That is, the transition of spots from rapid to slowly rotating latitudes and vice versa appears to be happening at a similar pace. Moreover, the minimum of $\avrelO$ occurs almost exactly at activity maximum ($\phi \approx 0.5$). Dividing this range into narrower bins in period or separating by temperature shows the same picture of a symmetric variation of $\avrelO$ with cycle phase. 

\item $\Prot = 22$-$28$ days: These stars show the strongest anti-correlation with the activity cycle, and also show the largest amplitude in both $\avrelI$ and $\avrelO$. Most clearly seen in Fig.~\ref{fig:eliptical}, there is a slight asymmetry in $\avrelO$ between the first and second halves of the activity cycle. During the rise of the cycle, the slope of $\avrelO$ is marginally steeper than during the declining phase of the cycle. Note however, that as with the faster rotators, the minimum in $\avrelO$ occurs at activity maximum.

For comparison we show the solar $\avrelI$ and $\avrelO$ in the inset in Fig.~\ref{fig:eliptical}, again based on the 11 year activity cycle. This shows a dramatically different picture of the change in spot rotation rotation with the activity cycle of the Sun, compared to the stars in our sample. The solar $\avrelO$ variation is also shown in blue in Fig.~\ref{fig:flip}, where we have scaled the differential rotation coefficients from \citet{Snodgrass1990} by a factor of $2.5$ for clarity. This illustrates that not only is the shape of the $\avrelO$ curve different, but it is also significantly out of phase with the stars in our sample stars. This produces the relation between $\avrelI$ and $\avrelO$ for the Sun seen in Fig.~\ref{fig:eliptical}.
\end{enumerate}

The characteristics of the behavior between the activity cycle and the changes in the rotation rate are summarized in Table~\ref{tab:chars}.
\begin{table*}
\caption{Characteristics of $\avrelI$ and $\avrelO$ in period ranges I, II, III, and IV, including the Sun for reference.}
\centering
\begin{tabular}{lccccccc}
Group & min($\avrelI$) [\%] & max($\avrelI$) [\%] & \makecell{$\avrelI$ \\phase at max.} & \makecell{$\avrelO$ \\amplitude [\%]} & \makecell{$\avrelO$ \\phase at min.} & $r_{av}$  \\
\hline
I, $\teff < 4200$K      & $-40.90\pm4.80$ & $52.05\pm7.39$  & $0.45$ 		& $0.59\pm0.13$ & $0.53\pm0.18$ 		& $-0.50$
\\
I, $\teff \geq 4200$K   & $-32.91\pm0.93$ & $45.02\pm1.77$  & $0.50$ 		& $0.33\pm0.04$ & $0.98\pm0.13$ 		& $0.76$ 
\\
II                      & $-37.09\pm1.01$ & $55.75\pm2.14$  & $0.50$ 		& $0.16\pm0.05$ & $0.57\pm0.31$ 		& $-0.45$ 
\\
III                     & $-41.19\pm0.89$ & $67.92\pm2.08$  & $0.50$ 		& $1.70\pm0.07$ & $0.49\pm0.04$ 		& $-0.95$ 
\\
IV                      & $-49.71\pm2.27$ & $91.10\pm6.03$  & $0.50$ 		& $2.86\pm0.25$ & $0.50\pm0.09$ 		& $-0.94$ 
\\
The Sun                 & $-70.58\pm5.03$ & $77.95\pm22.88$ & $0.50$		    & $1.24\pm0.09$	& $0.05$				& $0.15$ 
\\
\end{tabular}
\tablefoot{The values are extracted from sinusoidal fits to the data (dashed curves in Fig.~\ref{fig:flip}). For the fits, we use the same number of phase bins (20 bins) as shown in Fig.~\ref{fig:flip}.  For $\avrelI$ we show the minimum and maximum values, along with the  phase at activity maximum. The stars in range I are divided into two groups of cool stars ($\teff<4200$) and hot stars ($\teff \geq 4200$). The quantity $r_{av}$ is the correlation between the $\avrelI$ and $\avrelO$ of the stars in each subsample. For the solar case, the averages represent averages over multiple cycles and no sinusoidal fit was performed.}
\label{tab:chars}
\end{table*}

\section{Discussion}
Given the wide range of effects seen among stars of different rotation rates in our sample, we will separately discuss features seen among the fast rotators ($\Prot<15$ days) and slow rotators ($\Prot > 15$ days). The discussion of the latter is further broken down based on two separate assumptions: that we are observing the primary magnetic activity cycle, i.e., the equivalent to the solar 11 year cycle; or, given that the cycle periods of these stars is typically only a few years, that we are observing a faster secondary cycle. This division is motivated by the fact that multiple periodicities are seen in both solar and stellar time series, and because the \kepler time series are not long enough to detect stellar cycles longer than about 6 years.

In the following we make the assumption that the differential rotation profiles for the stars are solar-like, in the sense that the equatorial rotation rate is greater than at the poles. It is well established from both theory and numerical models of stellar convection \citep{Rudiger1989,Kueker2008,Miesch2009} that stars with rotation periods $\lesssim20$-$25$ days can establish a rapidly rotating equator compared to the polar regions (i.e. solar-like differential rotation). While strong Lorentz-force feedback from dynamo-generated magnetic fields may suppress the amplitude of the rotational shear in the fastest rotators (particularly for cool stars with deep convection zones), the sense of the rotation gradient should still be equatorward \citep{Brown2008,Augustson2012,Guerrero2013,Gastine2014}. For slowly-rotating stars with rotation periods longer than $\sim20$ days, the assumption of solar-like differential rotation is less reliable. Global convection simulations suggest that the rotation gradient may reverse to being poleward at the surface when the Rossby number exceeds unity \citep{Gastine2014,Featherstone2015,Karak2015}. However, the transition from solar to anti-solar differential rotation in convection simulations is rather abrupt and if it were occurring in our stellar sample, we would expect to see a dramatic change in behavior near a rotation period of about 25-30 days, depending on spectral type. Since we see no indication of such a dramatic transition, we will assume for the remainder of this section that all stars in our sample have a solar-like differential rotation. 

\subsection{Potential sources of sample contamination}
A potential source of noise is if the cycle period found by \citet{Reinhold2017} is not caused by an activity cycle. Instead, the observed variation of the flux such as or similar to that seen for \kicxample in Fig.~\ref{fig:cycle_example}, could be caused by a beating effect between two or more spots at slightly different latitudes, and thus with slightly different rotation rates. This is caused by one spot 'lapping' the other around the star. If in a given light curve, several beat periods are observed, this would be interpreted as a modulation in the photometric variability. With multiple spots on the stellar surface at once, several different spot configurations are possible but they may be roughly divided into two categories: 
(i) The difference in rotation rate between two active latitudes is small, leading to a long lap-time. If this lap-time is long compared to the typical spot lifetime, many spots will emerge and decay over the course of just a few lap-times. Since spot emergence is generally expected to be random in phase, the beat patterns would be similarly randomly distributed in time. We therefore anticipate the photometric variability to change stochastically, and such cases would likely fail the false-alarm probability test used by \citet{Reinhold2017}.
(ii) If the spots are long-lived compared to the lap-time, two spots may produce several beat periods. The observed change in the photometric variability would appear very periodic, and may therefore be falsely picked up as a cycle period. However, during one cycle (beat) the observed spot rotation rate as measured by the centroid of the power distribution in the frequency spectrum will not change. While the observed rotation rate might change between cycles (beats), when averaging over multiple cycles (beats) as is done here, the average change in the spot rotation rate will still be close to zero. We therefore expect such cases to contribute only to the number of stars with a very low degree of correlation, and not produce the observed changes in the spot rotation rates as seen in particular for the slow rotators. For the fast rotators however, this may be a source of contamination.

Furthermore, based on inferences from chromospheric and coronal emission, it has been suggested \citep{Berdyugina2005,Strassmeier2009} that especially fast rotators may, during high activity phases, become completely saturated with spots on the stellar surface. If spots cover a substantial fraction of the stellar surface the photometric variability may decrease, even as magnetic activity increases. For cases such as this the photometric variability is no longer a reliable way of measuring the activity cycle. This may account for the reduction in the amplitude of $\corr$ for rotation periods less than 11 days, as shown in Fig.~\ref{fig:corr}.

\subsection{Relationship between starspot rotation and activity among fast rotators}
The tendency of individual stars among the fast rotators, with a very low degree of correlation between $\relI$ and $\relO$, may in part be caused by contamination effects as discussed above. However, when averaging $\relO$ over multiple stars the correlation with the activity cycle becomes much more readily visible (see Fig.~\ref{fig:flip} frame I). One interpretation of this change in rotation rate is in terms of latitudinal migration of emerging starspots. If this is the case, then the low correlation for an individual star could be caused by a large scatter in emergence latitudes at a particular phase of the cycle. The mean emergence latitude may still migrate in latitude, but only when observing multiple stars will this become apparent. In contrast to the slower rotators in our sample, the fast rotating stars show a positive correlation between $\avrelI$ and $\avrelO$, with primarily the hotter stars showing this tendency. This effect is not unprecedented. \citet{Donahue1992} found that the star $\beta$ Com shows a clear increase of the rotation rate with activity, similar to what is observed here. On the other hand, as shown by Fig.~\ref{fig:corr} the cool stars ($\teff \lesssim 4200$K) in this period range show the opposite tendency, namely a negative correlation of $r_{av} = -0.50\pm0.13$ between $\avrelI$ and $\avrelO$. This suggests that a temperature dependence exists in this period range, between activity level and spot rotation, which is not present in the slower rotators in our sample.

One possible explanation for the low degree of correlation for the individual stars in this period range is that strong Lorentz-force feedback reduces the differential rotation in times of high activity. Since the equatorial region has much more mass and rotational inertia, this could lead to an increase in the rotation rate at mid-latitudes when the activity level is high. This could produce a change in the latitudes at which the spots emerge and a change in their rotation rate.

\subsection{Relationship between starspot rotation and activity among slow rotators}
\label{sec:hypos}
\begin{figure*}
   \centering
   \includegraphics[width=0.49\hsize]{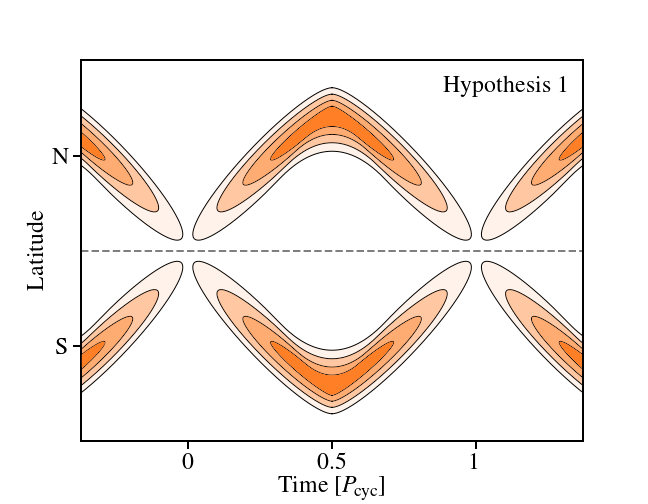}
   \includegraphics[width=0.49\hsize]{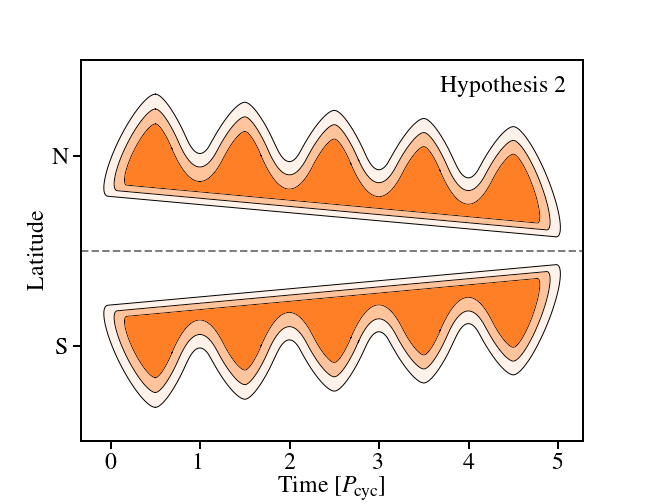}
      \caption{Qualitative illustrations of spot emergence patterns based on Hypothesis 1 (left) and Hypothesis 2 (right) as discussed in Section \ref{sec:hypos}. In both cases we assume a slowly rotating pole compared to the equator (similar to the Sun). Darker shades indicate higher activity levels (e.g. higher spot emergence rates). For the depiction of Hypothesis 2 we use the solar-like butterfly diagram simply as an example to describe the overall change in emergence latitude caused by the primary cycle. 
      }
          \label{fig:hypos}
\end{figure*}

The slow rotators in our sample would be expected to be those that show the greatest degree of similarity to the Sun. However, the changes in the spot rotation rate over the cycle show a dramatically different picture. The Sun shows changes in the spot rotation rate with a sharp transition between successive cycles and a large phase-lag between rotation minimum and activity maximum, leading on average to a positive correlation between the spot rotation and activity cycle. The stars in our sample on the other hand show a very symmetric variation in $\avrelO$, with a minimum at the phase of activity maximum. Assuming once more that the changes in starspot rotation rates are due to migration of active bands over differentially rotating latitudes, we suggest two different hypotheses based on two independent assumptions in order to explain the observations.

\subsubsection{Hypothesis \#1: Observed cycles are primary activity cycles}
The first hypothesis is based on the assumption that we are observing cycles that correspond to the solar $\sim11$ year cycle. In this case, assuming for the reasons described above that the differential rotation is solar-like, our findings indicate a butterfly diagram where the active regions emerge at low latitudes during the start of a cycle, at higher latitudes towards the peak of a cycle, and then at lower latitudes towards the end of a cycle. Figure~\ref{fig:hypos} shows an illustration of what the spot migration patterns on these stars might look like. Another possibility is that the butterfly diagrams of individual cycles look like those of the Sun (with equatorial propagation over the course of a cycle), but with a greater degree of overlap between successive cycles than the Sun exhibits. In this case at the start of a cycle the spots from the previous cycle could still be determining the inferred rotation rate. 

\subsubsection{Hypothesis \#2: Observed cycles are short secondary cycles}
Given the differences in $\avrelO$ between the Sun and stars, and also considering the relatively short cycle lengths of the stars, an alternative interpretation is that the observed cycles are shorter, secondary cycles and not equivalent to the solar 11 year cycle. Multiple stars have been shown to exhibit two distinct periods in magnetic activity \citep{See2016,Olah2016,Distefano2017}, one which is typically on decadal timescales and another much shorter on timescales of a few years. Since the \kepler mission lifetime limits us to cycle periods on the order of a few years, a bias should be expected toward these shorter cycles.

The Sun also exhibits a secondary cycle, the so-called quasi-biennial oscillation (QBO). The solar QBO is a  modulation of the radial magnetic field strength on $\sim1.5-3$ year timescales. This modulation is strongest at high latitudes, i.e., it affects primarily the top of the solar butterfly-diagram wings \citep{Ulrich2013}. In principle, when the QBO suppresses the primary cycle it could therefore cause fewer spots to emerge at high latitudes. On the other hand, when the QBO enhances the primary cycle, sunspot emergence would be enhanced at high latitudes. In terms of an average spot rotation period, this would lead to a modulation of the observed spot rotation rate which is out of phase with the change in activity level, leading to the observed anti-correlation. Figure~\ref{fig:hypos} shows an illustration of the modulation of the spot emergence latitude over a single primary cycle period. 

The solar QBO is observable in multiple different proxies including: $10.7$cm radio flux \citep{ValdesGalicia2007}, X-ray emission \citep{Antalova1994}, acoustic oscillation frequency modulations \citep{Fletcher2010}, as well as in the TSI and its variance used in this work. However, in our measurements of the solar $\relO$ this modulation is not readily visible. The solar 11 year cycle modulates the measured $\avrelO$ on the order of $1\%$, and so the QBO should be expected to produce even smaller variations in the spot rotation rate. In contrast, the amplitude of $\avrelO$ observed for the slow rotators in our sample is on the order of $3\%$. While the stars in our sample are likely much more active than the Sun, this large amplitude of the stellar $\avrelO$ is unexpected from a solar-like QBO.

In addition to the QBO’s, there is also the possibility of a non-axisymmetric quadrupole dynamo mode \citep[e.g.][]{Moss1995}. There is some observational support for such a mode existing on the Sun \citep{Berdyugina2003} and on other stars \citep{Tuominen2002,Berdyugina2005b}
in terms of active longitudes in each hemisphere separated by 180$^\circ$ in phase. The mode is cyclic, with a flip-flop transition between the two active
longitudes every 1.5 to 3 years in the case of the Sun \citep{Berdyugina2003}, similar to the period of the QBO’s.  The interaction between this mode and the primary (axisymmetric)
dipole mode of the star could also produce a butterfly diagram similar to that shown in the right panel of Fig.~\ref{fig:hypos}.

\section{Conclusions}
We measure the photometric variability as a proxy for magnetic activity, and star spot rotation rates as a function of time for $3093$ \kepler stars. These stars span a temperature range from $3200-6900$ and with rotation periods from $2-45$ days, and all have previously measured activity cycle periods from \citet{Reinhold2017} ranging from $0.5$ to $6$ years.

We find that the solar-like rotators tend to show a negative correlation between the activity level and rotation rate over the cycle period. For these stars the spots are at their slowest during activity maximum, and vice versa during activity minimum. The fast rotators ($\Prot<20$ days) on the other hand tend toward a decorrelation between activity and rotation. In terms of Rossby number\footnote{$Ro=1/(2\Omega_{\mathrm{rot}}\tau_{\mathrm{cz}})$, where the convective turnover time $\tau_{\mathrm{cz}}$ is computed based on \citet{Noyes1984a}.}, $Ro$, this transition occurs between $Ro\approx0.02$-$0.08$. This decorrelation can be caused by contamination primarily from misidentified cycle periods, or possibly by suppression of differential rotation by Lorentz-force feedback effects. However, in a narrow period range between $5-11$ days, stars with $\teff \gtrsim 4200$K show positive correlation, where the spot rotation maximum occurs during activity maximum. Stars cooler than this show the same negative correlation as the slower rotators. This indicates a temperature dependence on the relation between activity and spot rotation in this period range, something that is not seen for the slower rotators.

For the slow rotators the negative correlation leads to two different hypotheses, under the assumption that the change in rotation rate is caused by spot migration. The first hypothesis is that the spot rotation rates change because of a butterfly-like migration pattern caused by the primary dynamo cycle. The observed change in spot rotation rate would however lead to a dramatically different migration pattern, looking more like a boomerang than a butterfly wing, or would imply that the cycles overlap considerably more than on the Sun. The second hypothesis is that we are observing a shorter secondary cycle. In the Sun this is called the quasi-biennial oscillation, which could in principle modulate the emergence latitudes of spots, and thus produce changes in the spot rotation rate that are anti-correlated with the activity level.

Directly measuring the latitudes of starspots is still not possible for stars of broadly varying spectral types, as the ones in our sample. Measuring the variation of the spot rotation rates on the other hand provides an intriguing look at what phenomena might be present on active stars with short activity cycles. 

\begin{acknowledgements}
The authors would like to thank Timo Reinhold, Alexander Shapiro, and Gibor Basri for useful discussions. MBN and LG acknowledge support from NYUAD Institute grant G1502. This paper includes data collected by the \kepler mission. Funding for the Kepler mission is provided by the NASA Science Mission directorate. Data presented in this paper were obtained from the Mikulski Archive for Space Telescopes (MAST). STScI is operated by the Association of Universities for Research in Astronomy, Inc., under NASA contract NAS5-26555. Support for MAST for non-HST data is provided by the NASA Office of Space Science via grant NNX09AF08G and by other grants and contracts.
\end{acknowledgements}

\bibliographystyle{aa}
\bibliography{main.bib,more.bib}

\end{document}